\documentclass[reprint,aps]{revtex4-1}
\usepackage{amssymb}
\usepackage{amsmath}
\usepackage{graphicx}
\usepackage{multirow}
\usepackage{comment}
\usepackage{color}
\usepackage{xfrac}
\usepackage{threeparttable}
\usepackage{array}
\usepackage{wasysym}
\usepackage{subcaption}
\usepackage{floatrow}
\usepackage{float}
\floatstyle{plaintop}
\restylefloat{table}
\usepackage[colorlinks,linkcolor=blue,anchorcolor=black,citecolor=blue,urlcolor=black]{hyperref}
\usepackage[font=scriptsize,labelfont=bf]{caption}
\graphicspath{{./figures/}}

\begin{document}
\newcommand{\tny}[1]{\mbox{\tiny $#1$}}
\newcommand{\eqn}[1]{\mbox{Eq.\hspace{1pt}(\ref{#1})}}
\newcommand{\eqs}[2]{\mbox{Eq.\hspace{1pt}(\ref{#1}--\ref{#2})}}
\newcommand{\eqsu}[2]{\mbox{Eqs.\hspace{1pt}(\ref{#1},\ref{#2})}}
\newcommand{\eqtn}[2]{\begin{equation} \label{#1} #2 \end{equation}}
\newcommand{\func}[1]{#1 \left[ \rho \right] }
\newcommand{\mfunc}[2]{#1_{#2} \left[ \rho \right] }
\newcommand{\mmfunc}[3]{#1_{#2} \left[ #3 \right] }
\newcommand{\mmmfunc}[4]{#1_{#2}^{#3} \left[ #4 \right] }
\newcommand{\pot}[1]{v_{\rm #1}}
\newcommand{\spot}[2]{v_{\rm #1}^{#2}}
\newcommand{\code}[1]{\texttt{#1}}
\newcommand{\beq}{\begin{equation}}
\newcommand{\eeq}{\end{equation}}
\newcommand{\bea}{\begin{eqnarray}}
\newcommand{\eea}{\end{eqnarray}}

\def\brpppp{{\mathbf{r}^{\prime\prime\prime\prime}}}
\def\brppp{{\mathbf{r}^{\prime\prime\prime}}}
\def\brpp{{\mathbf{r}^{\prime\prime}}}
\def\brp{{\mathbf{r}^{\prime}}}
\def\bzp{{\mathbf{z}^{\prime}}}
\def\bxp{{\mathbf{x}^{\prime}}}
\def\tp{{{t}^{\prime}}}
\def\tpp{{{t}^{\prime\prime}}}
\def\tppp{{{t}^{\prime\prime\prime}}}

\def\tbr{{\tilde{\mathbf{r}}}}
\def\bk{{\mathbf{k}}}
\def\bq{{\mathbf{q}}}
\def\br{{\mathbf{r}}}
\def\bz{{\mathbf{z}}}
\def\bx{{\mathbf{x}}}
\def\bR{{\mathbf{R}}}
\def\bM{{\mathbf{M}}}
\def\bP{{\mathbf{P}}}
\def\bT{{\mathbf{T}}}
\def\bK{{\mathbf{K}}}
\def\bA{{\mathbf{A}}}
\def\bB{{\mathbf{B}}}
\def\bG{{\mathbf{G}}}
\def\bX{{\mathbf{X}}}
\def\bY{{\mathbf{Y}}}
\def\bP{{\mathbf{P}}}
\def\bI{{\mathbf{I}}}
\def\d{{\mathrm{d}}}
\def\rhor{{\rho({\bf r})}}
\def\rhorp{{\rho({\bf r}^{\prime})}}
\def\rhoi{{\rho_I}}
\def\rhoii{{\rho_{II}}}
\def\rhoj{{\rho_J}}
\def\rhoir{{\rho_I({\bf r})}}
\def\rhoiir{{\rho_{II}({\bf r})}}
\def\rhojr{{\rho_J({\bf r})}}
\def\rhoirp{{\rho_I({\bf r}^{\prime})}}
\def\rhojrp{{\rho_J({\bf r}^{\prime})}}
\def\sumi{{\sum_I^{N_S}}}
\def\sumj{{\sum_J^{N_S}}}
\def\im{{\operatorname{Im}}}

\def\etal{{\it et al.}}
\def\vdw{{van der Waals}}
\def\vw{{von Weizs\"{a}cker}}
\def\qe{{\sc Quantum ESPRESSO}}
\def\se{{Schr\"{o}dinger equation}}
\def\ses{{Schr\"{o}dinger equations}}
\def\bnabla{{\boldsymbol{\nabla}}}
\def\bchi{{\boldsymbol\chi}}
\def\bLambda{{\boldsymbol\Lambda}}
\def\bDelta{{\boldsymbol\Delta}}

\title{Machine Learning Electronic Structure Methods Based On The One-Electron Reduced Density Matrix}

\author{Xuecheng Shao}
\email{xuecheng.shao@rutgers.edu}
\affiliation{Department of Chemistry, Rutgers University, Newark, NJ 07102, USA}

\author{Lukas Paetow}
\affiliation{Department of Chemistry, Rutgers University, Newark, NJ 07102, USA}

\author{Mark E. Tuckerman}
\email{mark.tuckerman@nyu.edu}
\affiliation{Department of Chemistry, New York University, New York, NY 10003, USA}
\affiliation{Courant Institute of Mathematical Science, New York University, New York, NY 10003, USA}
\affiliation{Simons Center for Computational Physical Chemistry, New York University, New York, NY 10003, USA}
\affiliation{NYU-ECNU Center for Computational Chemistry at NYU Shanghai, Shanghai 200062, China}

\author{Michele Pavanello}
\email{m.pavanello@rutgers.edu}
\affiliation{Department of Physics, Rutgers University, Newark, NJ 07102, USA}
\affiliation{Department of Chemistry, Rutgers University, Newark, NJ 07102, USA}

\begin{abstract}
    The theorems of density functional theory (DFT) and reduced density matrix functional theory (RDMFT) establish a bijective map between the external potential of a many-body system and its electron density or one-particle reduced density matrix.  Building on this foundation, we show that machine learning can be used to generate surrogate electronic structure methods. In particular, we generate surrogates of local and hybrid DFT as well as Hartree-Fock and full configuration interaction theories for systems ranging from small molecules such as water to more complex compounds like propanol and benzene. The surrogate models use the one-electron reduced density matrix as the central quantity to be learned.  From the predicted density matrices, we show that either standard quantum chemistry or a second machine-learning model can be used to compute molecular observables, energies, and atomic forces. We show that the surrogate models can generate essentially anything that a standard electronic structure method can, ranging from band gaps and Kohn-Sham orbitals to energy-conserving ab-initio molecular dynamics simulations and IR spectra, which account anharmonicity and thermal effects, without the need of computationally expensive algorithms such as self-consistent field theory. The algorithms developed here are packaged in an efficient and easy to use Python code, QMLearn, accessible by the broader community on popular platforms.  
\end{abstract}
\maketitle

\section{Introduction}
Computational models are routinely employed to predict molecular and material properties in lieu of or prior to performing costly experiments. They are also used to explain the complex electron and nuclear dynamics that underlie experimental observations. \cite{cart2008}.
When these computational strategies require the evaluation of the electronic structure of a system, they often become the computational bottleneck, lengthening the time to solution.  Consequently, an important and timely goal is the development of approaches capable of providing the electronic structure of complex systems at reduced computational cost \cite{Koopman_2019,Bannwarth_2019,JKD+19,Qiao_2020} or even bypassing electronic structure calculations altogether.  This article focuses on achieving the latter by leveraging the power of machine learning (ML).

The standard use of ML methods is to target single quantities of interest, which are learned in terms of a few descriptors. Examples of this are predictions of the electronic energy (including those that apply the concept of ``delta'' learning)  \cite{chen2020ground,christensen2021orbnet,welborn2018transferability,dick2020machine}, dipole moments, and polarizabilities \cite{Wilkins_2019,Willatt_2019}, to name a few.  Such a modus operandi is not ideal. Should the target be a quantity for which the model has not been trained, then a new model needs to be trained to predict this quantity. A relevant example is the computation of IR spectra, where both the spectral line positions as well as the intensities are needed. A typical ML model that learns the potential energy surface of a molecular system or a material can only predict the spectral line positions through a molecular dynamics simulation followed by analysis of the velocity autocorrelation function \cite{marx_book}. However, in order to predict intensities, the autocorrelation function of the dipole moment is needed. The usual procedure to set up two ML models -- one for the energy surface and one for the dipole moment -- is time consuming and ultimately avoidable. 

\begin{table*}[htp]
    \caption{\label{tab:method}Brief summary of the ML methods employed in this work: $\gamma$-learning and $\gamma+\delta$-learning. $E$ and $F$ stand for electronic energy and atomic forces, respectively. The row ``Predicted'' introduces the nomenclature of the quantities predicted by the ML methods. See text for further details. rLR stands for regularized linear regression.}
    \begin{tabular}{l@{~~~}l@{~~~}l@{~~~~}l@{~~~~}l}
        \hline
        & Model       & Features          & Targets                 & Prediction \\[5pt]
        $\gamma$-learning        & \eqn{vgkrr} &$\hat v$ & $\hat \gamma$ & $\hat \gamma^p$ \\[5pt]
        $\gamma+\delta$-learning & rLR & $\hat \gamma^p$ & $\hat \gamma$, $E$, $F$ & $\hat \gamma^p+\Delta \hat \gamma^p$, $E_p$, $F_p$\\
        \hline
    \end{tabular}
\end{table*}

An ideal ML method should learn and learn from quantities that are dense in the amount of information they hold. The most general and ideal ML model would learn the many-body electronic wavefunction. From the wavefunction, one could predict the potential energy surface, the dipole moment and any other quantity of interest. ML models of many-body wavefunctions \cite{Hermann_2020,li2022fermionic} are becoming competitive against other accurate wavefunction solvers, such as Quantum Monte Carlo. However, the complexity involved in both computations and training data sizes \cite{Westermayr_2021,Husch_2021} hampers their broad applicability. Luckily, rigorous and bijective maps from DFT  \cite{Hohenberg1964PR,levy1988} and Reduced Density Matrix Functional Theory (RDMFT) \cite{gilbert1975hohenberg,Donnelly_1978,Valone_1980,Levy_1979} can be exploited to shift the focus from the many-body wavefunction to other, related quantities such as the electron density, $\rho(\br)$, or the $N$-electron reduced density matrix ($N$-rdm), $\hat\gamma_N$.

The electron density can be learned in terms of local atom-centered descriptors \cite{Willatt_2019,lewis2021learning,mahmoud2020learning}. However, the model densities typically deviate from the target by a few percent, which is not sufficient for their use in electronic structure methods, where the self-consistent electron densities need to be converged to within much tighter thresholds. Exploiting the Hohenberg and Kohn theorems \cite{Hohenberg1964PR}, the electron density can be used as the target quantity in ML models where external potentials serve as features \cite{Brockherde_2017,Bogojeski_2020}. This recovers model electron densities that, even if not completely accurate, can still be successfully used as a feature to accurately learn the energy and forces of methods such as DFT and coupled cluster \cite{christensen2021orbnet,Bogojeski_2020,Yao_2016}.

Focusing on the 1-rdm instead of the electron density has several advantages, e.g., the ability to deliver expectation values of any one-electron operator, including non-multiplicative operators such as the kinetic energy, the exchange energy, and the corresponding non-local (Hartree-Fock) potential. Additionally, much as can be done for the electron density, formal functionals of the 1-rdm can be learned, such as that of the electronic energy or corresponding atomic forces. Therefore, models for the 1-rdm should significantly extend the scope of ML models compared to those that learn the density alone \cite{Wetherell_2020}.

In this work, we take on the challenge of learning 1-rdms to such an accuracy that the predicted 1-rdm are essentially indistinguishable from those delivered by standard electronic structure software. We achieve this aim by representing external potentials and target 1-rdms in terms of their matrix elements over Gaussian Type Orbitals (GTOs) and devising an efficient generator of training sets. This allows us to deliver ``surrogate electronic structure methods'' that use the 1-rdm to predict derived, useful quantities (energy, forces, band gaps, orbitals) that are equally accurate and useful as those computed by standard electronic structure software. 

In the following sections, we present examples of calculations performed using various surrogate electronic structure methods, ranging from DFT to full configuration interaction, after introducing the main algorithms. Our proof of concept consists of seven molecules from small to medium-sized, rigid and floppy, which we have chosen to demonstrate the uniqueness and novelty of our method. To the best of our knowledge, our approach is the first to date to successfully create fully-fledged electronic structure surrogates.

\begin{table*}[htbp]
    \caption{Benchmark study for LDA$^{\rm ML}$, the surrogate electronic structure method for LDA KSDFT. RMSDs of predicted energy, magnitude of the atomic forces, magnitude of the dipole moment vector, and the non-interacting kinetic energy, along a 10 ps AIMD trajectory sampled every 100 fs (100 test structures in total) at 300 K. LDA$^{\rm ML}[\gamma^{p}]$ and LDA$^{\rm ML}[\hat\gamma^{p}+\Delta\hat\gamma^{p}]$ compute all quantities from the predicted 1-rdms. $N_{vib}$ is the number of vibrational degrees of freedom in the molecule.}
	\label{tab:w_l}
	\setlength{\tabcolsep}{3mm}{
	\begin{tabular}{c|l|rrrrrrr}
	\hline
	\multicolumn{2}{c|}{System}                                                                                                 & H$_2$O & CO$_2$ & NH$_3$ & CH$_3$OH$^{*}$ & C$_6$H$_6$ & 1-propanol$^{*}$ & 2-propanol$^{*}$ \\
	\hline
        \multicolumn{2}{c|}{$N_{vib}$}                                                                                          & 3      & 3      & 6      & 12       & 30         & 30         & 30         \\
	\hline
        \multicolumn{2}{c|}{Training set size ($N_{sample}$)}                                                                   & 27     & 27     & 216    & 5184     & 13824      & 41472      & 41472      \\
	\hline
        \multirow{3}{*}{\parbox{2.5cm}{\centering Energy (meV/$N_{vib}$)}}    & LDA$^{\rm ML}[\gamma^{p}]$                      & 0.006  & 0.030  & 0.006    & 0.018      & 0.021    & 0.237       & 0.092 \\
                                                                              & LDA$^{\rm ML}[\gamma^{p}+\Delta\gamma^{p}]$     & 0.005  & 0.029  & 0.006    & 0.005      & 0.008    & 0.039       & 0.011 \\
                                                                              & LDA$^{\rm ML}$                                  & 0.337  & 0.407  & 0.027    & 0.042      & 0.021    & 0.228       & 0.186 \\ \hline
    \multirow{3}{*}{\parbox{2.5cm}{\centering Force (meV/\AA/$N_{vib}$)}}     & LDA$^{\rm ML}[\gamma^{p}]$                      & 22.17  & 37.00  & 12.57    & 10.08      & 5.47     & 7.80        & 5.30  \\
                                                                              & LDA$^{\rm ML}[\gamma^{p}+\Delta\gamma^{p}]$     & 7.67   & 3.50   & 1.08     & 3.81       & 1.52     & 3.40        & 1.49  \\
                                                                              & LDA$^{\rm ML}$                                  & 1.26   & 1.09   & 0.11     & 0.25       & 0.08     & 1.67        & 0.59  \\ \hline
    \multirow{2}{*}{\parbox{2.5cm}{\centering Dipole ($10^{-3}$D/$N_{vib}$)}} & LDA$^{\rm ML}[\gamma^{p}]$                      & 0.57   & 0.25   & 0.19     & 0.13       & 0.13     & 0.35        & 0.35  \\
                                                                              & LDA$^{\rm ML}[\gamma^{p}+\Delta\gamma^{p}]$     & 0.05   & 0.06   & 0.02     & 0.05       & 0.02     & 0.16        & 0.08  \\ \hline
    \multirow{2}{*}{\parbox{2.5cm}{\centering Kinetic Energy (meV/$N_{vib}$)}}            & LDA$^{\rm ML}[\gamma^{p}]$          & 4.03   & 8.87   & 4.05     & 5.28       & 1.62     & 7.83        & 5.97  \\
                                                                              & LDA$^{\rm ML}[\gamma^{p}+\Delta\gamma^{p}]$     & 1.56   & 0.85   & 0.25     & 0.69       & 0.50     & 1.57        & 1.45  \\ \hline

	\end{tabular}}
	\flushleft{\footnotesize{$^*$ generate samples for 3 stable conformers.}}
\end{table*}

\section{Details of the surrogate electronic structure methods}

\subsection{Learning rigorous maps from DFT and RDMFT}
We aim at learning rigorous maps from DFT \cite{Hohenberg1964PR} and RDMFT  \cite{gilbert1975hohenberg} linking the 1-rdm (the full matrix for RDMFT or just the real-space diagonal elements for DFT) with virtually any ground state property given as the expectation value of any operator.  Specifically,  the following two maps are considered (dropping subscripts for notational ease),
\begin{align}
    \label{gv}
    \tag{map 1}
    \hat v \to \hat \gamma, \\
    \label{gE}
    \tag{map 2}
    \hat\gamma \to E, F, \langle \hat O \rangle,
\end{align}
where $\hat v$ is the external potential, $E$ and $F$ are the electronic energy and the corresponding atomic forces, respectively, and $\langle \hat O \rangle$ is the expectation value of the operator $\hat O$. We call the ML procedure for \ref{gv} $\gamma$-learning, and that for \ref{gE}  $\gamma+\delta$-learning. We note that when KSDFT methods are considered, the KS 1-rdm, $\hat\gamma_s$, is targeted instead of $\hat\gamma_1$.

The utility of a ML model for \ref{gv} is evident, as the computationally costly steps in electronic structure solvers (such as the SCF procedure or even more complex algorithms for  post-HF methods) are replaced by the ML model. For similar reasons, the utility of learning \ref{gE} is also evident. When considering mean field methods, \ref{gE} can either be learned or directly computed from predicted 1-rdms. We will consider these options and show that they both lead to equally successful outcomes (vide infra).

In this work, we represent 1-rdms and external potentials in terms of GTOs.  GTOs are useful and convenient. For example, expectation values are simply computed, e.g., the  non-interacting kinetic energy, $T_s[\hat \gamma]=-\frac{1}{2}{\rm Tr} \left[ \nabla_{\brp}^2 \gamma(\br,\brp) \right]={\rm Tr} \left[ \hat \gamma \hat t \right]$ (where $\nabla_{\brp}$ is the gradient with respect to the $\brp$ variable, $t_{\mu\nu}=-\frac{1}{2}\langle \mu | \nabla^2 | \nu \rangle$, where $\mu$ and $\nu$ are GTO indices). Another important positive consequence of using GTOs is the possibility to work in an internal reference frame, providing a straightforward framework for dealing with the rotational and translational degrees of freedom.
For ML models, this has been a significant challenge \cite{Marcos_2016,Goodfellow-et-al-2016,Willatt_2019} 

\subsection{Learning \ref{gv}: $\gamma$-learning}
Inspired by Brockherde \etal\ \cite{Brockherde_2017}, we learn \ref{gv} by supervised ML exploiting a kernel ridge regression (KRR), 
\eqtn{vgkrr}{\hat \gamma [\hat v] = \sum_i^{N_{sample}} \hat \beta_i K(\hat v_i,\hat v).}
In the above equation $K(\hat v_i,\hat v_j)={\rm Tr}[\hat v_i\hat v_j]$, $\{\hat v_i\}$ is a training set of size $N_{sample}$ of external potentials and $\hat \beta_i$ are KRR coefficients which, in this case, are matrices of leading dimension $N_{AO}$, the size of the chosen GTO basis set. 
The $\hat \beta_i$ coefficients are determined by KRR through the standard method of inversion of the regularized kernel matrix leading to (in matrix notation) $\hat \beta_i = \sum_j \left[ \mathbf{K} + \lambda \mathbf{I} \right]^{-1}_{ij}\hat \gamma_j$, where $\mathbf{K}_{ij} = K(\hat v_i,\hat v_j)$, $\mathbf{I}$ is the identity matrix, $\lambda$ a regularization hyperparameter, and $\{\hat \gamma_j\}$ the target 1-rdms from the training set. The 1-rdms predicted by $\gamma$-learning are denoted by $\hat\gamma_p$.

\subsection{Learning \ref{gE}: $\gamma+\delta$-learning}

While in $\gamma$-learning external potentials serve as features and the 1-rdms are the targets, in $\gamma+\delta$-learning, due to the shifted focus to learning functionals of the 1-rdm, the features are the matrix elements of the 1-rdm. Targets can be the 1-rdm itself (producing predicted 1-rdms of higher accuracy, hereafter denoted by $\hat\gamma^p+\Delta\hat\gamma^p$), the electronic energies and associated nuclear forces (hereafter denoted by $E_p$ and $F_p$), or, in principle, any other expectation value with respect to the ground state wavefunction. Another difference among the two learning steps are the types of regression used. As we have seen, KRR is employed for $\gamma$-learning, while a regularized linear regression is used for $\gamma+\delta$-learning. A summary of $\gamma$- and $\gamma+\delta$-learning and the adopted nomenclature for the associated predicted quantities is given in Table \ref{tab:method}.

Once the ML models are formulated, the next step is to develop a strategy for training them, which is the subject of the next section.

\subsection{Training and benchmark tests}

$\gamma$- and $\gamma+\delta$-learning require carefully generated training sets. A set of molecular geometries that samples a desired configuration space needs to be formulated. In a second step, for $\gamma$-learning external potential/1-rdm pairs of matrices over a GTO basis for each geometry in the training set are needed. For $\gamma+\delta$-learning, the electronic energy, atomic forces, and any additional molecular property are needed. Separate training sets need to be generated for each molecule and for each electronic structure method targeted.

Key to the success of this effort is to limit the training set to the smallest number of elements possible. Our guiding principle is to train the ML models, carrying out only a fraction of the simulations needed to run a standard AIMD simulation with the target electronic structure method. 

\begin{figure*}[htbp]
    \includegraphics[width=\textwidth]{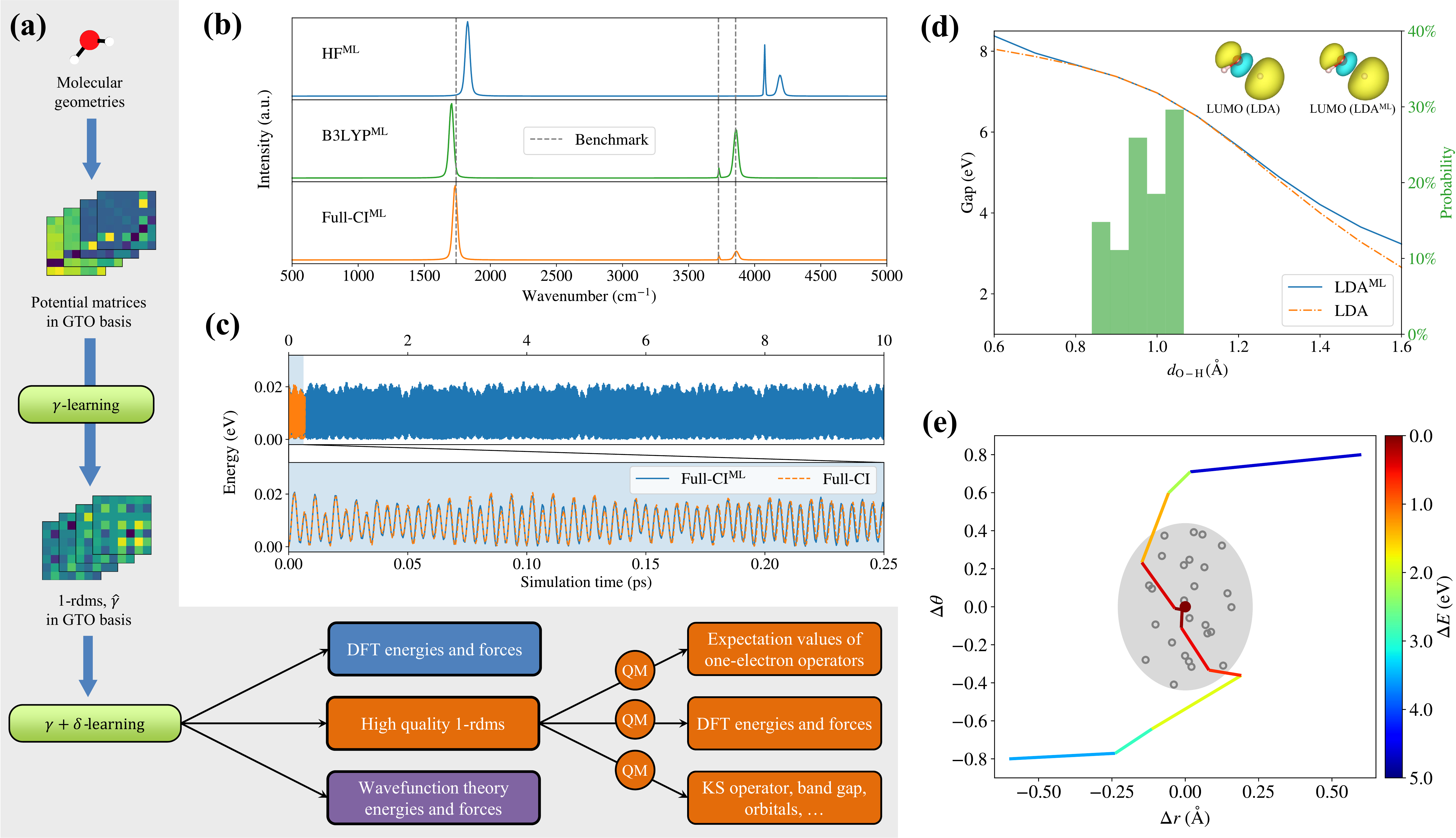}
    \caption{The performance of surrogate electronic structure methods (superscript ML) for water is presented. (a) The ML methods used to make predictions are described. For each molecular geometry, the matrix elements of the external potential (e.g. electron-nuclear attraction potential) with respect to a GTO basis set are computed. $\gamma$-learning uses 1-rdms as targets in the GTO basis via \eqn{vgkrr}, exploiting \ref{gv}. $\gamma + \delta$-learning targets electronic energies and forces from any desired electronic structure method, and further refines the 1-rdm itself, exploiting \ref{gE}. (b) The predicted IR spectrum of water is compared to CCSD(T) Car-Parrinello molecular dynamics vibrational frequencies which are taken to be our benchmark, shown with dashed lines in the figure. (c) The electronic energy along one AIMD trajectory with initial random velocities consistent with a temperature of 300 K in the NVE ensemble is shown. (d) The HOMO-LUMO band gap as a function of the OH distance ($d_{O-H}$) is plotted, along with the associated LUMO orbital for $d_{O-H} = 1.6$\AA. The training set structures are indicated by a green histogram. (e) The geometry relaxations of H$_2$O molecules are carried out with Full-CI$^{\rm ML}$. The difference in OH distances ($\Delta r$ in \AA) and the difference in HOH angle ($\Delta \theta$ in radians) from the equilibrium angle are shown. The training set structures are indicated by empty grey circles and the equilibrium geometry is represented by a dark red dot. The light grey area indicates the configuration space covered by the training set.}
\label{fig:water}
\end{figure*}

Such a tall order can be achieved using a sampling method based on vibrational normal mode analysis. For each molecule we (1) carry out a normal mode analysis at the equilibrium geometry, and (2) generate $N_{sample}$ geometries randomly displaced from the equilibrium geometry along each normal mode according to a normal (Gaussian) distribution of variance $\sigma_i^2 = 2k_BT N_{atoms} /[\Omega_i^2 N_{vib}\left(1-2/9N_{vib}\right)^3]$, where $T$ is a target temperature, $N_{vib}$ is the number of vibrational degrees of freedom, and $N_{atoms}$ is the number of atoms, and $\Omega_i$ is the vibrational frequency of mode $i$. In principle, sampling $N_{vib}$ modes with $N_{s}$ points for each mode would lead to $N_{sample}=\left(N_{s}\right)^{N_{vib}}$ total sampled points, which would be unattainable. However, after some testing, we noticed that $N_{sample}\propto N_{vib}^3$ is more than sufficient to achieve a deviation of less than 1 meV for the predicted energy per vibrational degree of freedom. For floppy molecules we carry out samplings from several initial geometries that reflect the number of stable conformers.

As shown in Table \ref{tab:w_l}, the training set size, $N_{sample}$, is 27 (or $N_{sample}=N_{vib}^3$) for molecules with $N_{vib}=3$, and it grows to 13824 for benzene which has $N_{vib}=30$ (i.e., $N_{sample}=(0.8\cdot N_{vib})^3$). $N_{sample}$ then triples for methanol, 1- and 2-propanol due to the need to generate samples for the three most stable conformers (found by rotating the H atom of the OH group around the axis given by the CO bond). 

Supplementary Figure S1 shows histograms of the electronic energies corresponding to the sampled geometries for benzene in comparison to an AIMD simulation. The sampling method captures, within a margin of error, the electronic energy distribution with a much reduced number of energy evaluations compared to the AIMD simulation. In Figure S2 we show that the distribution of geometries in each normal mode are also correctly recovered. The distributions are given in the energy scaled displacement coordinates (i.e., it is expected that all distributions share the same energy-scaled variance, $\sigma^2=\sigma_i^2\Omega_i^2$).

Table \ref{tab:w_l} reports results for a surrogate of KSDFT within the local-density approximation (LDA). The parent method is indicated by LDA and the surrogate model by LDA$^{\rm ML}$. We compute and predict quantities along a 10 ps AIMD for several molecular systems, from small and rigid (such as water) to medium size and floppy (such as 1- and 2-propanol). 

Energies, forces, dipoles, and non-interacting kinetic energies can either be predicted through separate regressions using $\gamma + \delta$-learning, or calculated from the predicted 1-rdm obtained from $\gamma$-learning (i.e., with $\hat{\gamma}^p$) or $\gamma + \delta$-learning (i.e., with $\hat{\gamma}^p + \Delta\hat{\gamma}^p$).
In Table \ref{tab:w_l}, dipole moments and non-interacting kinetic energies are only computed from the predicted 1-rdms. When employing the most accurate model, dipole deviations are at most 10$^{-4}$ D  per vibrational degree of freedom, and non-interacting kinetic energies deviate by a few meV per vibrational degree of freedom. 

Table \ref{tab:w_l} also shows that energies and forces computed by the 1-rdm from  $\gamma+\delta$-learning (indicated by the predicted 1-rdm, $\hat \gamma^p+\Delta\hat \gamma^p$), substantially improve the result from $\gamma$-learning. While the electronic energies do not improve much, the computed atomic forces improve more, deviating by only 1-8 meV/\AA\ per vibrational degree of freedom. Such a deviation is acceptable for AIMD and geometry optimization. As expected, computed expectation values, i.e., dipoles and non-interacting kinetic energy are also improved when they are evaluated with $\hat \gamma^p+\Delta\hat \gamma^p$ rather than just with $\hat \gamma^p$.

\subsection{The QMLearn software and workflow}
The methods developed in this work are collected in the all-Python QMLearn software which is freely available from GitLab \cite{qmlearn} and easily installed through \code{pip install qmlearn}. QMLearn is composed of the following classes: (1) a database collecting the training sets; (2) QM engines (we use PySCF although other engines are also supported) capable of generating the training sets and matrix elements over the GTOs as well as the needed infrastructure to compute energies and atomic forces; (3) a structure handler (we use Atomistic Simulation Environment (ASE) \cite{ASE}). ASE is used to handle molecular geometry, including driving molecular dynamics simulations.; and (4) ML modules such as scikit-learn \cite{scikit-learn} or Tensorflow \cite{tensorflow} (the current version of QMLearn supports only scikit-learn).

Additional information is available in the supplementary information document.

\section{Showcasing several surrogate electronic structure methods for water}
Figure \ref{fig:water} showcases the performance of several surrogate electronic structure methods for the water molecule. Water is a small molecule, with only 3 normal modes of vibration, it is feasible to run full configuration interaction (Full-CI) simulations with the 6-31G* basis set and train the surrogate Full-CI$^{\rm ML}$ over a training set of only 27 structures. In addition to Full-CI, we also developed surrogates for Hartree-Fock, KSDFT within the LDA, and KSDFT using the hybrid B3LYP exchange-correlation functional. We tested the surrogate models in such common tasks as the computation of band gaps as a function of varying molecular geometry, the shape of HOMO and LUMO orbitals, NVE AIMD simulations, the prediction of IR spectra, and geometry relaxation (see the caption of Figure \ref{fig:water} for additional details as well as the supplementary materials for additional information on how the IR spectrum, orbitals, and band gaps of water were obtained).

For an AIMD simulation, Full-CI$^{\rm ML}$ is stable for tens of picoseconds and produces energies that are indistinguishable from those of a standard Full-CI simulation (which could be run for a shorter simulation time and is found to be five orders of magnitude slower than Full-CI$^{\rm ML}$). When the IR spectrum is calculated using KSDFT surrogates, it deviates somewhat from the coupled-cluster singles and doubles with perturbative triples (CCSD(T)) benchmark of Ref. \citenum{wathelet1998vibrational}. As expected, when the IR spectrum is calculated using Full-CI$^{\rm ML}$, it matches the benchmark's vibrational frequencies.

The band gap from the surrogate LDA KSDFT method, LDA$^{\rm ML}$, follows closely the parent LDA result; the orbitals do as well and are essentially indistinguishable from LDA. A similarly accurate result is found for geometry relaxations, where the equilibrium geometry is found starting from structures that are far away from the training set.

We therefore conclude that the surrogate models considered for the water molecule deliver a predicted electronic structure that is essentially indistinguishable from the parent, standard electronic structure method even for molecular geometries that are well outside the configuration space spanned by the training set.

\section{Ab-initio dynamics and IR spectra}
\begin{figure}[htbp]
	\captionsetup[subfigure]{skip=-10pt,margin={-15pt,0pt}}
	\begin{subfigure}{0.90\textwidth}
		\subcaption{}
		\includegraphics[width=\textwidth]{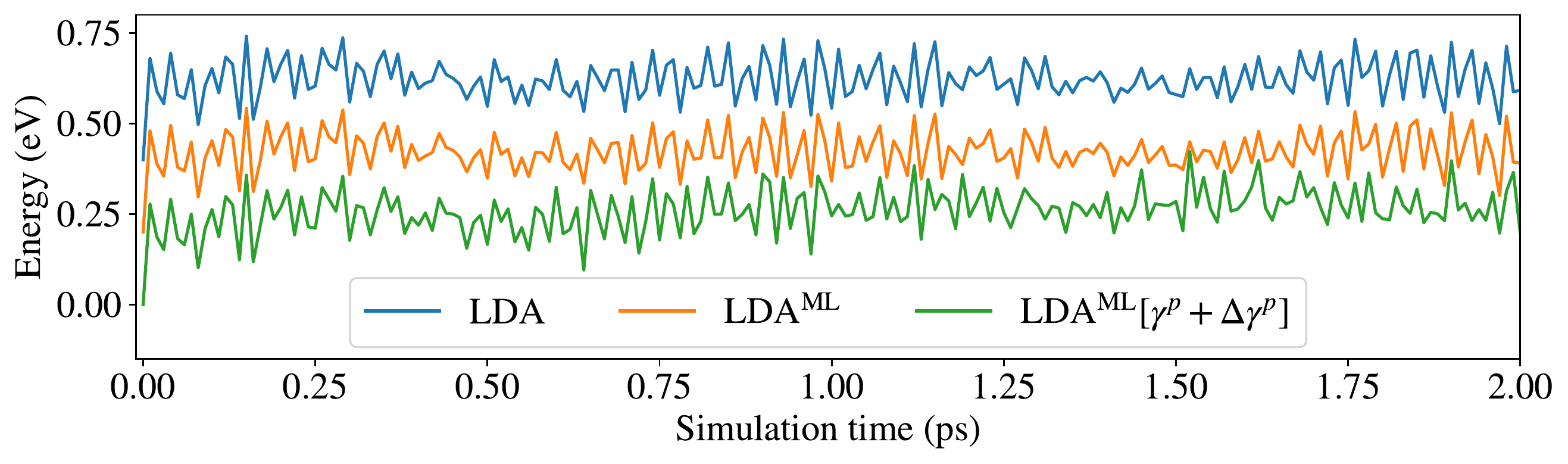}\label{fig:benz_md}
	\end{subfigure}
	\begin{subfigure}{0.90\textwidth}
		\subcaption{}
		\includegraphics[width=\textwidth]{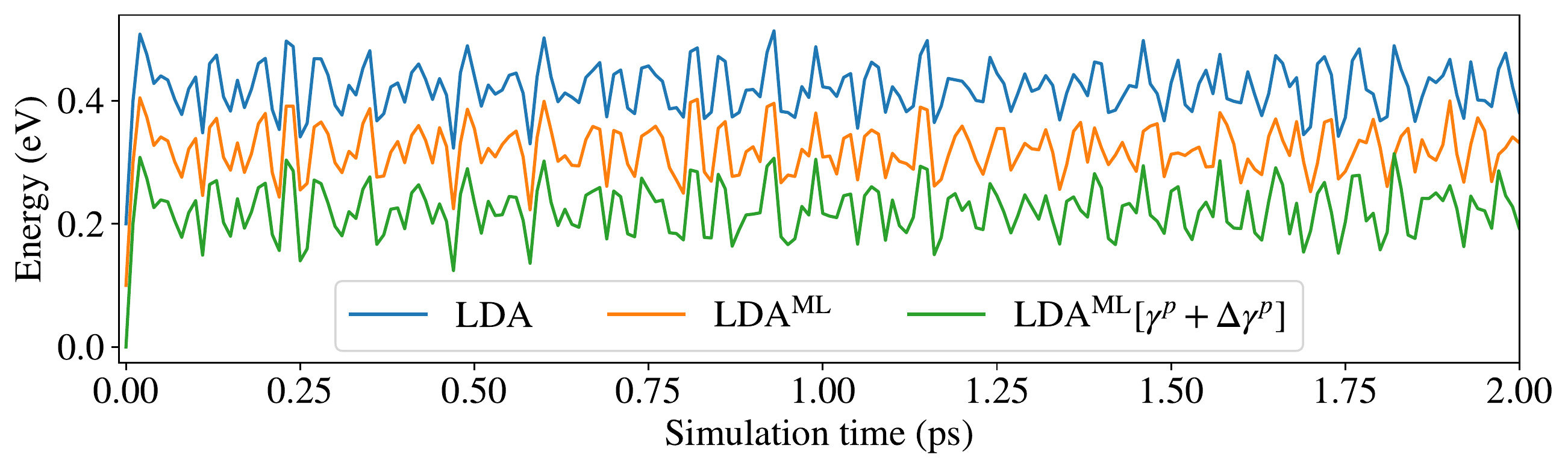}\label{fig:prop_md}
	\end{subfigure}
    \caption{\label{fig:md} Electronic energy values along three independent AIMD trajectories for (a) benzene and (b) 1-propanol carried out in the NVE ensemble with initial velocities corresponding to an instantaneous temperature of 300 K. The three trajectories share the same initial conditions but are run with three distinct methods: LDA$^{\rm ML}$, LDA$^{\rm ML}[\hat \gamma^p+\Delta\hat \gamma^p]$ (which uses forces and energies computed from the predicted 1-rdm) and the LDA benchmark trajectory. We only report energies from snapshots taken every 10 fs of simulation time.}
\end{figure}

A more stringent test of the robustness of the surrogate electronic structure methods is the generation of fully deterministic energy-conserving dynamics. Figure \ref{fig:md} reports the electronic energies along NVE trajectories (the equilibrium geometry was used as initial geometry and the same random initial velocities were used for the methods reported) at an instantaneous temperature of 300 K for benzene and 1-propanol.  Two flavors of the LDA$^{\rm ML}$ surrogate model are considered: LDA$^{\rm ML}$ and LDA$^{\rm ML}[\hat \gamma^p+\Delta\hat \gamma^p]$. The latter only predicts 1-rdms (the $\hat \gamma^p+\Delta\hat \gamma^p$) and computes energies and forces from them. 

Using LDA$^{\rm ML}[\hat \gamma^p+\Delta\hat \gamma^p]$ and associated forces or the predicted energies and forces for LDA$^{\rm ML}$ leads to essentially equivalent results. Specifically we see that trajectories from the LDA surrogates follow closely the parent LDA result up to a simulation time of 1 ps, after which they begin to slightly deviate. This behavior is in line with the RMSDs presented in Table \ref{tab:w_l}.

The surrogate electronic structure methods are able to predict, for example, molecular dipole moments along a AIMD trajectory directly from the predicted 1-rdms (i.e., the $\hat \gamma^p+\Delta\hat \gamma^p$). This can be exploited to compute anharmonic and temperature-dependent spectra of molecules \cite{marx_book}.

\begin{figure}[htbp]
	\captionsetup[subfigure]{skip=-10pt,margin={-15pt,0pt}}
	\begin{subfigure}{0.90\textwidth}
		\subcaption{}
		\includegraphics[width=\textwidth]{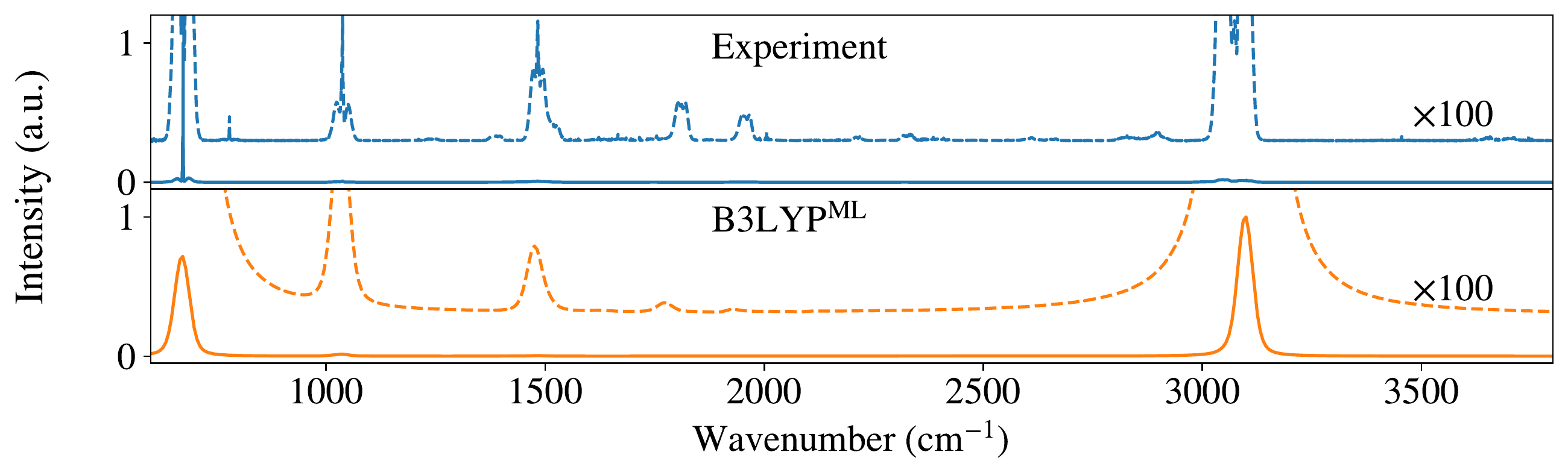}\label{fig:benz_ir}
		\subcaption{}
		\includegraphics[width=\textwidth]{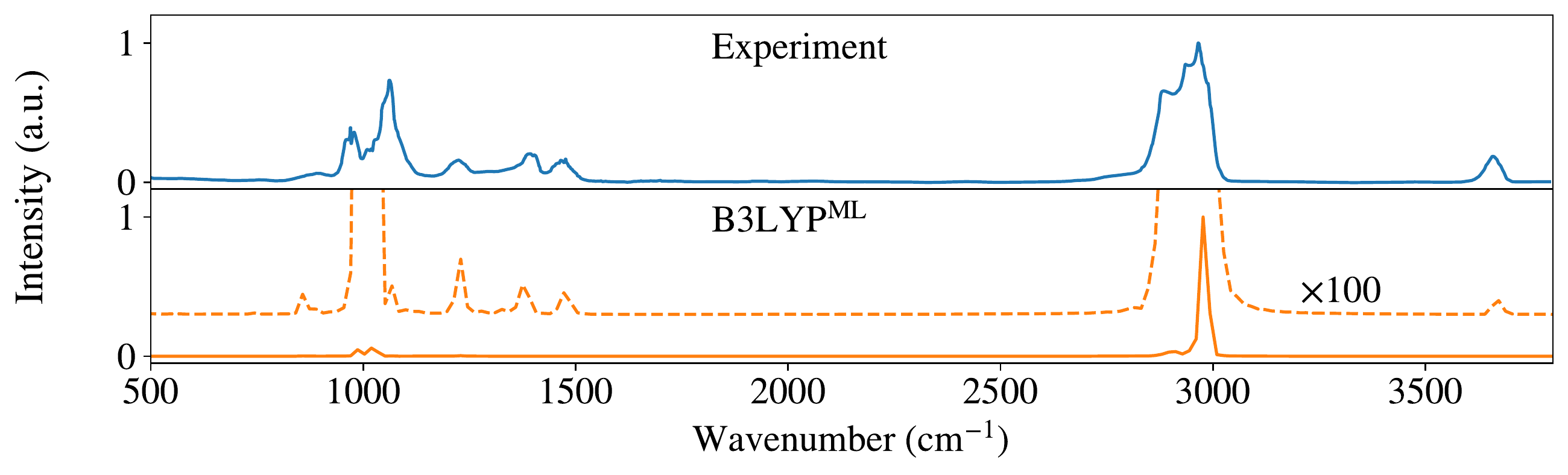}\label{fig:prop_ir}
		\subcaption{}
		\includegraphics[width=\textwidth]{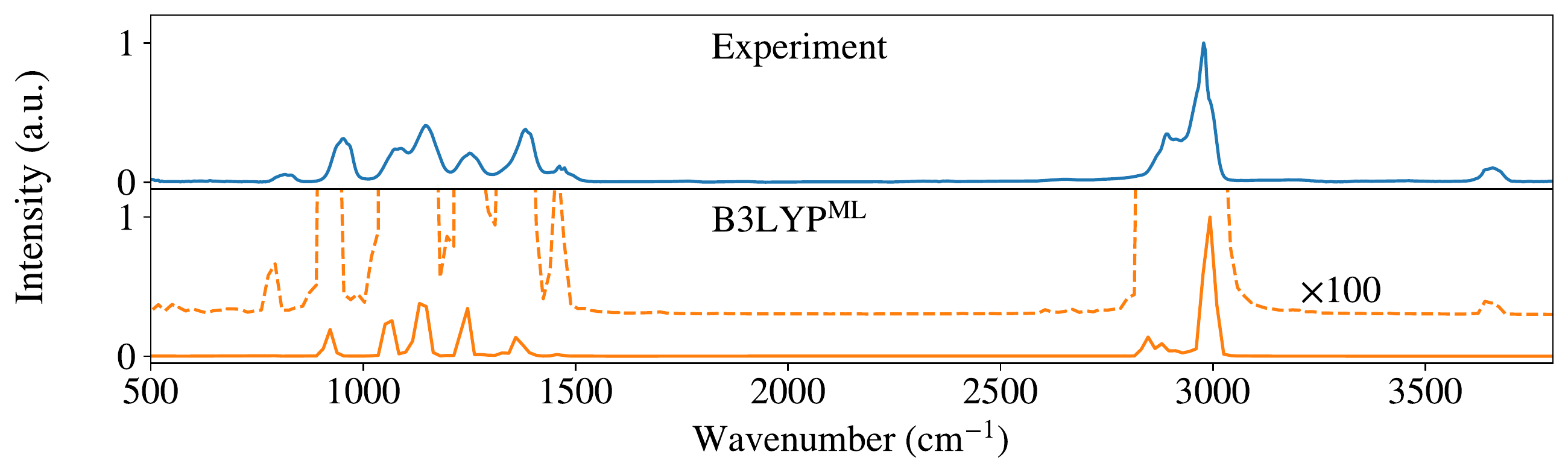}\label{fig:prop2_ir}
	\end{subfigure}
    \caption{\label{fig:ir}Predicted gas phase IR spectra of (a) benzene, (b)  1-propanol and (c) 2-propanol with the B3LYP$^{\rm ML}$ surrogate model. Gas phase experimental spectra from the NIST Standard Reference Database Number 69 \cite{NISTwebbook} are reproduced for comparison. A vibrational scaling factor of 0.97 was applied to the computed spectra. For ease of visualization, we also feature magnified intensities by a factor of 100.}
\end{figure}

Figure \ref{fig:water}a already reports the IR spectrum of a water molecule. Moving on to larger and more floppy molecular systems, in Figure \ref{fig:ir} we report the gas phase IR spectra of benzene, 1- and 2-propanol computed by the surrogate B3LYP$^{\rm ML}$ method. Given the large number of vibrational normal modes, for each of the three molecules we first run a 10 ps NVT dynamics at 300 K with B3LYP$^{\rm ML}$. We then selected five snapshots for benzene and twelve snapshots for 1- and 2-propanol (sampled between 5 and 10 ps of the trajectory) and run additional 2 ps NVE trajectories computing the dipole moment at each step. IR spectra are computed by Fourier transformation of the dipole autocorrelation function (additional details in the supplementary information document). Comparison to experimental spectra shows that the peak positions are well reproduced  and that the peak intensities follow the correct trend. Given the floppy nature of these molecules (especially for 1-propanol), it is conceivable that a larger number of snapshots and longer trajectories would provide a closer comparison to the experiment. However, such study goes beyond the scope of the present work.

\section{Predicted HOMO-LUMO gap and orbitals}

\begin{figure}[htbp]
	\begin{subfigure}{.48\textwidth}
		\subcaption{}
		\includegraphics[width=\textwidth]{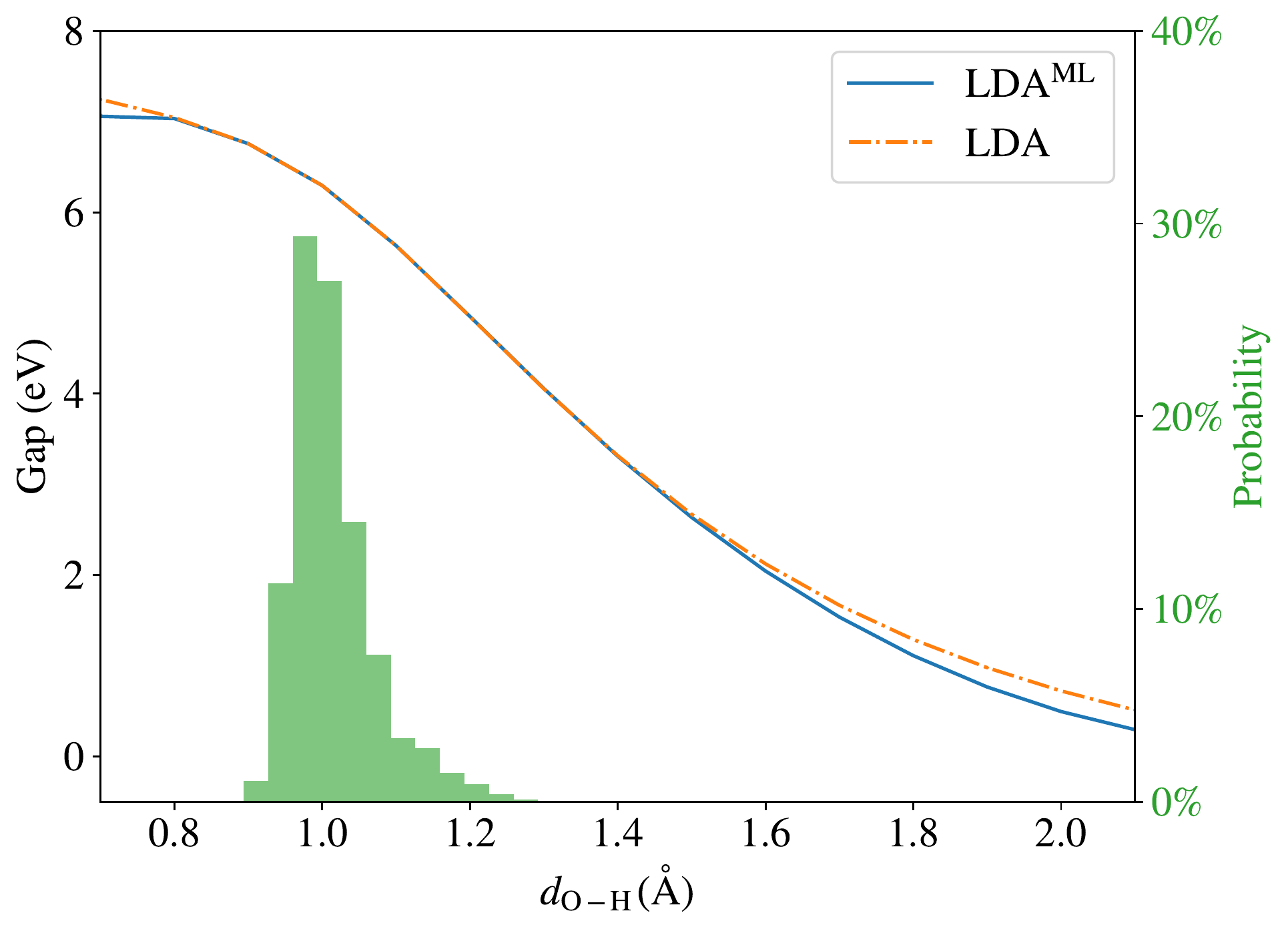}\label{fig:meth_gap}
	\end{subfigure}
	\begin{subfigure}{.48\textwidth}
		\subcaption{}
		\includegraphics[width=\textwidth]{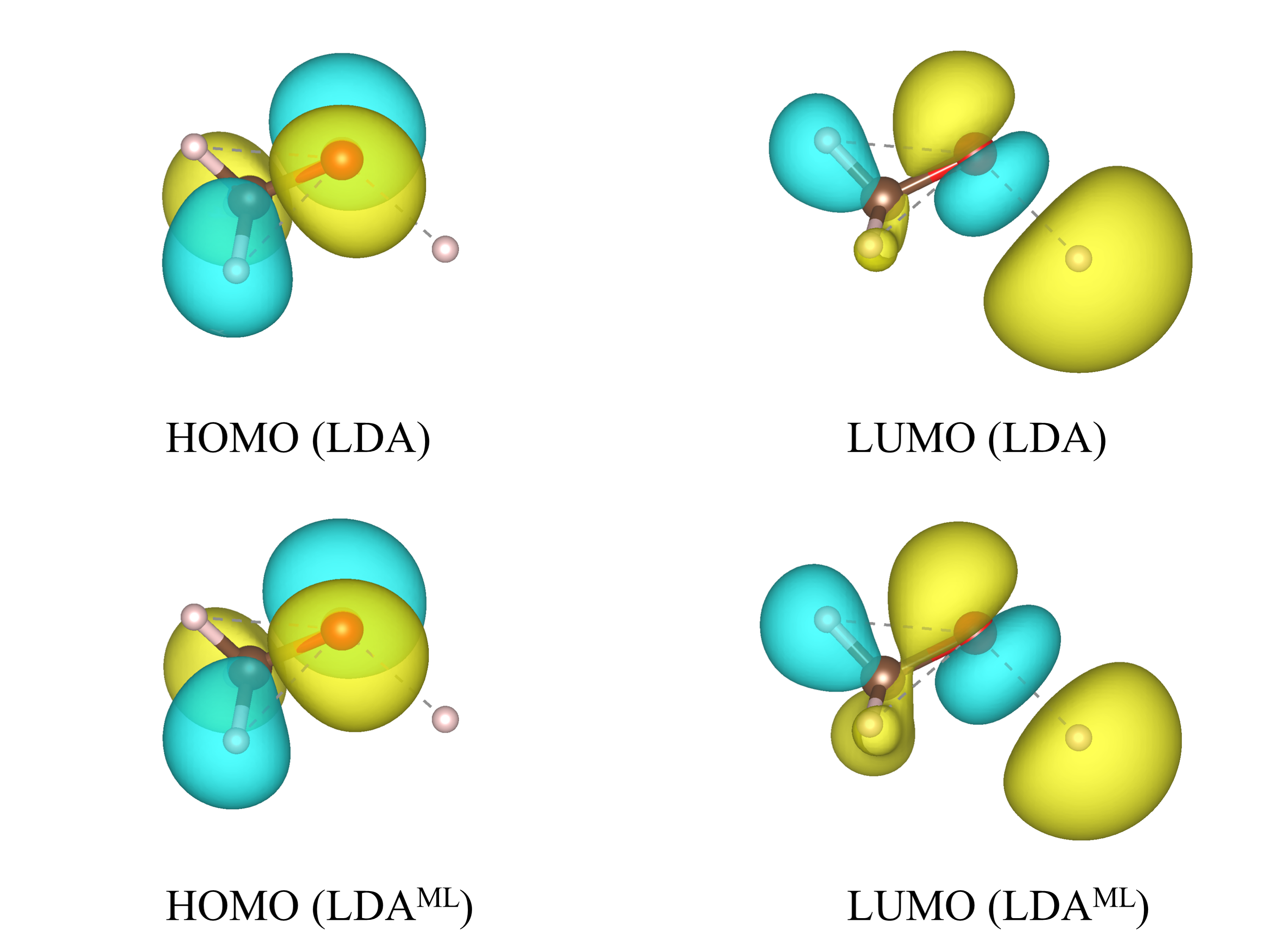}\label{fig:meth_orb}
	\end{subfigure}
    \caption{\label{fig:meth} (a) The highest occupied molecular orbital (HOMO)--lowest unoccupied molecular orbital (LUMO) energy gap as a function of OH distance ($d_{O-H}$) for methanol. The green histogram is the probability distribution of the OH distance, $d_{O-H}$, for the training set. (b) HOMO and LUMO orbital isosurfaces from LDA$^{\rm ML}$ and LDA at the OH distance of $d_{O-H}=2.0$ \AA.}
\end{figure}

\begin{figure}[htbp]
	\begin{subfigure}{.48\textwidth}
		\subcaption{}
		\includegraphics[width=\textwidth]{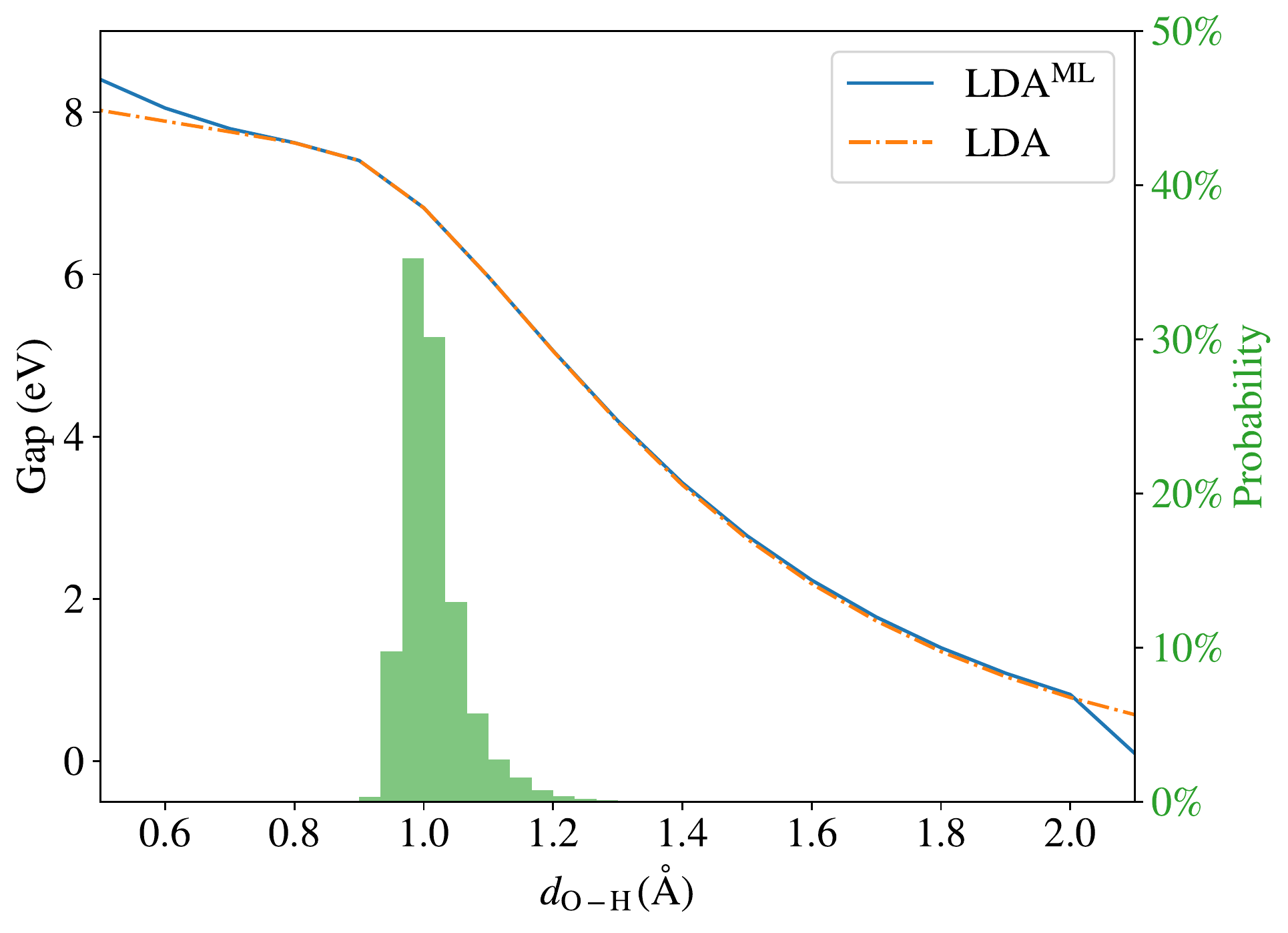}\label{fig:prop_gap}
	\end{subfigure}
	\begin{subfigure}{.48\textwidth}
		\subcaption{}
		\includegraphics[width=\textwidth]{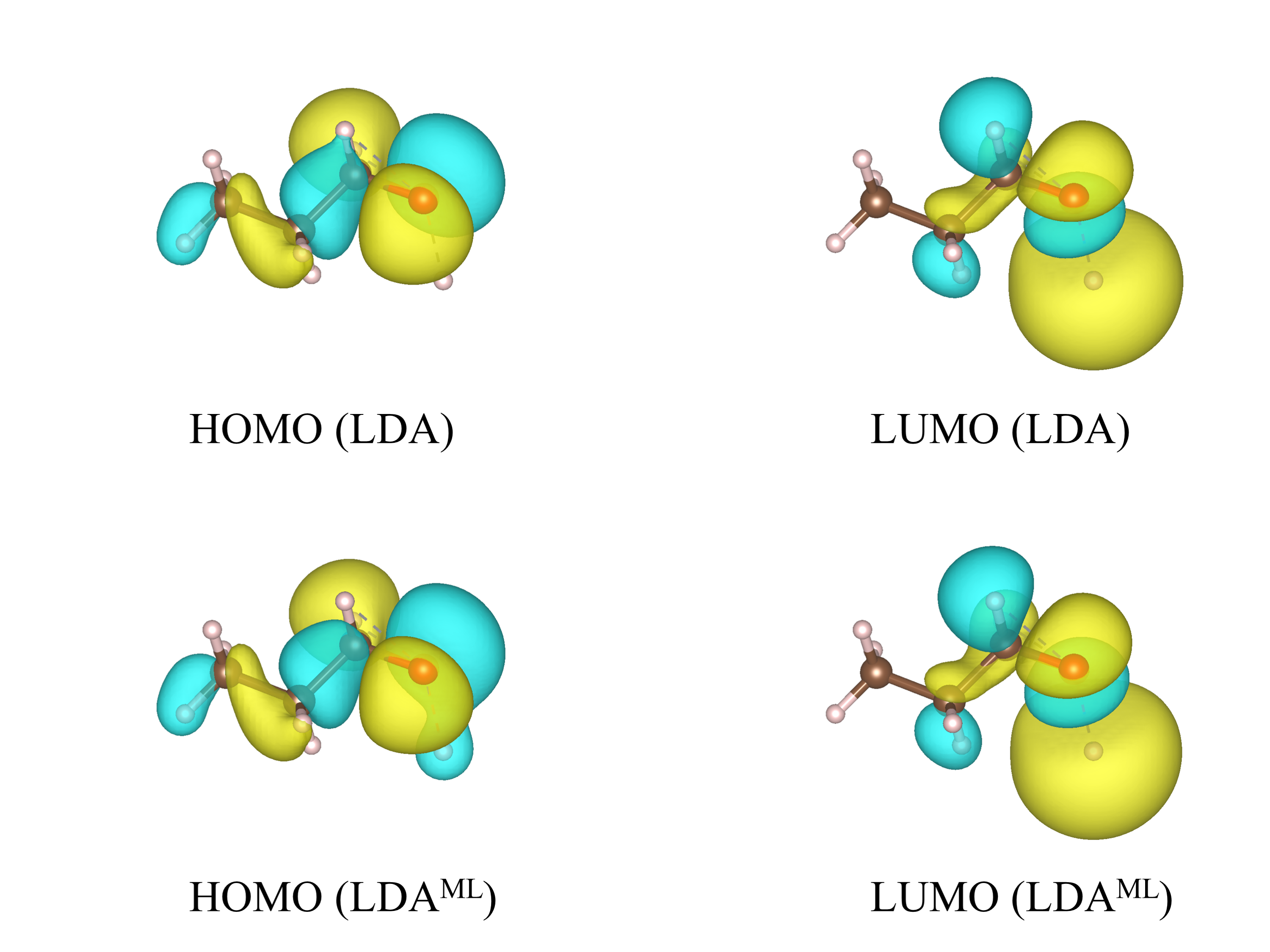}\label{fig:prop_orb}
	\end{subfigure}
    \caption{\label{fig:prop} HOMO-LUMO gap and orbital isosurfaces for 1-propanol. Other details are the same as the caption of Figure \ref{fig:meth}.}
\end{figure}

KSDFT methods provide the ability to calculate band gaps and generate HOMO and LUMO orbitals. This is useful for a variety of purposes, including the interpretation of photophysics and reactivity \cite{Houk_2021}. We therefore found it important to showcase the ability of the surrogate models to predict band gaps and orbitals for molecules other than water and for geometries near and far from the configuration space spanned by the training set. In addition to the already discussed Figure \ref{fig:water}d, Figures \ref{fig:meth} and \ref{fig:prop} consider methanol and 1-propanol where the OH bond is artificially shortened and stretched in 0.7--2.1 \AA and 0.5--2.1 \AA, respectively. The predicted LDA$^{\rm ML}$ HOMO-LUMO gap is given in panels (a) of the figures and is overall in excellent agreement with the LDA result for both molecules. Similar to what we witnessed for water, LDA vs LDA$^{\rm ML}$ gaps start to deviate only for geometries far away from the training set, see green histogram of OH bond distances from the training set reproduced in panels (a).

To appreciate the power of these surrogate models, we also reproduced with isosurface plots the predicted HOMO and LUMO orbitals for the most stretched configuration (OH bond length of 2 \AA). The results are similar to what we presented for water, i.e., they show that the orbitals from surrogate models are correct even for geometries very far away from the configuration space spanned by the training set.

\section{Conclusions}
We developed surrogate electronic structure methods based on machine learning of the one-electron reduced density matrix, or 1-rdm. When given a target molecule, the surrogate methods provide the same information and predictions as traditional electronic structure software.

We showcased DFT, HF and post-HF surrogates for rigid molecules such as water and benzene and for flexible molecules such as 1- and 2-propanol. First, our models learned the 1-rdm of these systems as a function of the external potential. Then, rigorous maps were machine learned from DFT and RDMFT to predict the 1-rdm to energy and 1-rdm to atomic forces maps.

Our surrogate methods robustly predict geometry optimizations, ab-initio dynamics, and IR spectra from the molecular dipole moment. Because of their versatility, the surrogate methods predict not only structure and dynamics but also expectation values of one-electron operators and Kohn-Sham orbitals. We predicted HOMO and LUMO orbitals and energy gaps for several molecules, with results nearly indistinguishable from traditional methods even for geometries far away from the configuration space sampled by the training sets.

This work is a proof of concept showing that surrogate electronic structure methods can replace conventional electronic structure methods for most computational chemistry tasks. However, the extending the algorithm to higher-order rdms would enable the computation of expectation values of two-electron operators, energies and forces and will be considered in the future. Condensed phase systems and larger molecules are also targeted for future development. Large molecules can be approached by combining the 1-rdms of separate molecular fragments learned from smaller molecules. For molecular condensed phase systems, density embedding can be employed as a divide-and-conquer tool resulting in a computationally low-scaling method. 

\section{Computational Details}
We  generate training sets and to compute all the needed matrix elements over the AOs with PySCF \cite{PYSCF}. KSDFT calculations carried out with the LDA xc functional use the parametrization from Perdew and Zunger \cite{Perdew1981}.
Unless otherwise stated, we use the cc-pVTZ basis set for all systems except benzene, 1- and 2-propanol for which we use the 6-31G* basis set.

ASE \cite{ASE} is employed to drive all ab-initio molecular dynamics (AIMD) simulations and geometry optimizations. Regressions are carried out with sci-kit learn \cite{scikit-learn}. 
Additional details (software, data availability, and other details) are available in the supplementary information document.

\section*{Acknowledgements}
This material is based upon work supported by the National Science Foundation under Grants No.\ CHE-1553993, CHE-2154760, OAC-1931473, and Petroleum Research Fund grant No.\ 62555-ND6.  M.E.T. acknowledges support from the Camille and Henry Dreyfus Foundation grant no. ML-22-146.
We thank the Office of Advanced Research Computing at Rutgers for providing access to the Amarel cluster.
\bibliography{paper}
\end{document}